\documentclass[numbers]{elsarticle}
\usepackage{latexsym}
\usepackage{natbib}
\usepackage[latin2]{inputenc}
\makeatletter
\def\ps@pprintTitle{%
	\let\@oddhead\@empty
	\let\@evenhead\@empty
	\def\@oddfoot{\centerline{\thepage}}%
	\let\@evenfoot\@oddfoot}
\makeatother

\begin{document}

\title{The Budapest Reference Connectome Server v2.0}
	
\author[p]{Balázs Szalkai\corref{cor2}}
\ead{szalkai@pitgroup.org}
\author[p]{Csaba Kerepesi\corref{cor2}}
\ead{kerepesi@pitgroup.org}
\author[p]{Bálint Varga\corref{cor2}}
\ead{balorkany@pitgroup.org}
\author[p,u]{Vince Grolmusz\corref{cor1}}
\ead{grolmusz@pitgroup.org}
\cortext[cor1]{Corresponding author}
\cortext[cor2]{Joint first authors}
\address[p]{PIT Bioinformatics Group, Eötvös University, H-1117 Budapest, Hungary}
\address[u]{Uratim Ltd., H-1118 Budapest, Hungary}

\date{}

\begin{abstract}
{Motivation:} The connectomes of different human brains are pairwise distinct: we cannot talk about an abstract "graph of the brain". Two typical connectomes, however, have quite a few common graph edges that may describe the same connections between the same cortical areas. 

{Results:} The Budapest Reference Connectome Server v2.0 generates the common edges of the connectomes of 96 distinct cortexes, each with 1015 vertices, computed from 96 MRI data sets of the Human Connectome Project. The user may set numerous parameters for the identification and filtering of common edges, and the graphs are downloadable in both csv and GraphML formats; both formats carry the anatomical annotations of the vertices, generated by the Freesurfer program. The resulting consensus graph is also automatically visualized in a 3D rotating brain model on the website. 
The consensus graphs, generated with various parameter settings, can be used as reference connectomes based on different, independent MRI images, therefore they may serve as reduced-error, low-noise, robust graph representations of the human brain.

{Availability:} http://connectome.pitgroup.org 
\end{abstract}

\maketitle

\section{Introduction} 

 Several large-scale projects for brain--mapping are being executed \cite{Johansen-Berg2013,McNab2013}, but the neuron-scale graph of the human brain, where the nodes are the neurons, and two neurons are connected by an edge if they are joined through a synapse, is out of reach today \cite{Yook2013}. The difficulties come from the number of the neurons to be mapped, and also from the lack of the high-throughput methods for mapping their connections. The neuron-scale graphs were constructed only for very simple organisms with a very small number of neurons \cite{Chklovskii2010,deLange2014,Towlson2013} or for just small cortical areas of more complex organisms \cite{Gilbert2013,Anderson2011}. 

The application of magnetic resonance imaging (MRI) offers numerous methods for mapping physical and functional connections between subdivided anatomical areas of the brain (called  "Regions of Interests", ROIs), each consisting of millions of neurons. The vertices are the ROIs, and two ROIs are connected by an edge if connections are detected between them by an MRI-based method. This method can either be diffusion MRI imaging, depicting the Brownian motion of water molecules in axons, consequently, mapping the axons between different cortical areas; or functional MRI (fMRI) imaging, depicting brain areas of elevated blood flow while the subject rests or performs different mental tasks. 

\begin{table*}[t!]
	\centering 
	{\small
		\begin{tabular}{ | l | l | }
		\hline
		Label & Description   \\ \hline
		id\_node1 & the numerical ID of the first vertex of the edge\\ 
		id\_node2 & the numerical ID of the second vertex of the edge\\
		name\_node1 & the anatomical name of node 1 \\
		name\_node2 & the anatomical name of node 2 \\
		parent\_id\_node1 & the ID of the parent region of node 1 on the 83-region atlas\\
		parent\_id\_node2 & the ID of the parent region of node 2 on the 83-region  atlas\\
		parent\_name\_node1 &  the name of the parent region of node 1 on the 83-region atlas\\
		parent\_name\_node2 &  the name of the parent region of node 2 on the 83-region atlas\\
		minimum\_edge\_confidence & the number of the graphs in which the edge is contained \\
		median & the median of the weights of the same edge in different graphs\\
		average & the average of the weights of the same edge in different graphs\\
		\hline
	\end{tabular}}
	\smallskip
	\caption{The column labels of the result file in csv format. The 83-region atlas refers to the atlas of the FreeSurfer tool.}
\end{table*}

In this note we present a web-server, which, starting from the diffusion MRI data published as a result of the Human Connectome Project \cite{McNab2013}, compiles differently parametrized reference graphs from the common edges of the graphs describing 96 different 1015-vertex graphs of 96 human subjects. Additionally, a default, single graph, the Budapest Reference Connectome v2.0 is also presented in two downloadable formats.

The resulting graphs may be used for identifying more robust, more error-free connections between the cortical areas, represented by ROIs: for example, in the default reference graph (i.e., the Budapest Reference Connectome v2.0), if an edge is present then it is present in at least 14 different source graphs. In general, one may set the "Minimum edge confidence" to value $k$ anywhere between $k=1$ (where an edge is included if it is present in at least one source graph) through $k=96$ (where an edge is present in the reference graph if it can be found in all the 96 source graphs).

Therefore, the resulting graphs contain {\it common, consensus} edges (i.e., Fig. 1) originated from multiple graphs with user-specified parameters, computed from the diffusion MRI data of different subjects.

\begin{figure}[h!]
	\centering
	\includegraphics[width=95mm]{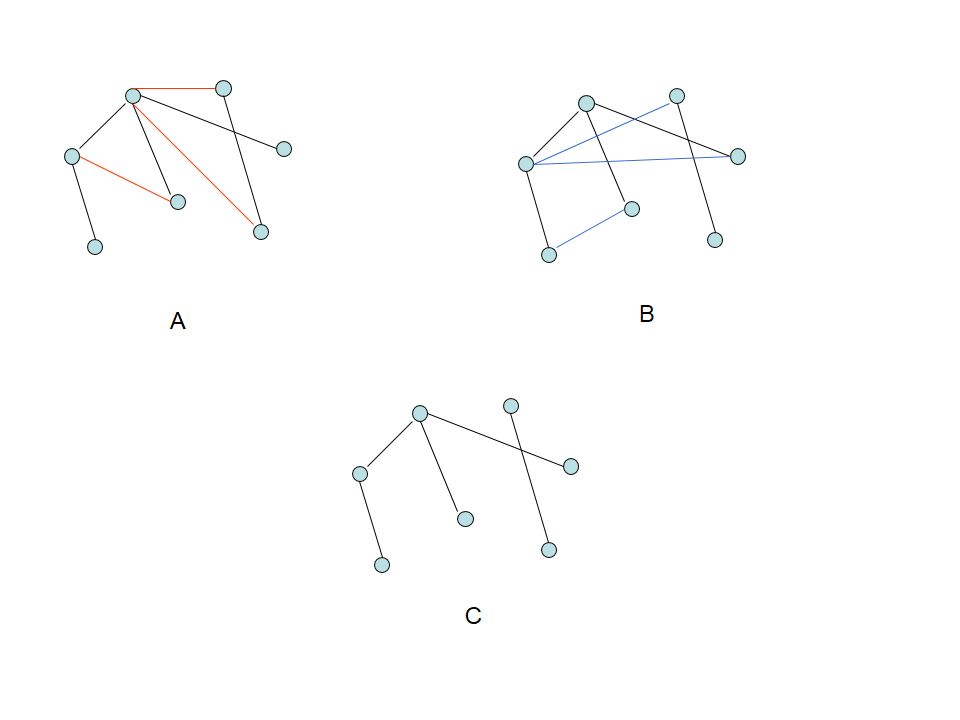}
	\caption{The black edges of graphs A and B are common edges; they form graph C, the consensus graph. }
\end{figure}

Version 2.0 of the Budapest Reference Connectome Server is described here in detail. Choosing Version 1.0 is also possible on the website: Version 1.0 applies the source data from the classical article of \cite{Hagmann2008} describing six connectomes of five subjects, each with 998 vertices. Version 1.0 of the webserver computes the consensus edges, with some parameter options, from those six graphs only.

By filtering edges with very few occurrences or those with small weights, one may get a connectome with more reliable edges and weights than in the case of any single dataset in the input. Therefore, we may get robust edges and weights in the consensus graphs generated by the server.

\section{Results and Discussion}

The webserver is available at http://connectome.pitgroup.org. The default, canonical ``Budapest Reference Connectome v2.0'' can be downloaded by simply hitting the "Download graph" button without changing the default options. This default graph has 1015 vertices, 8507 edges.

The following options can be set after choosing the "Show options" button:
\begin{itemize}
	
\item[(i)] Version 1.0 or Version 2.0. The default choice is 2.0, using the graphs of 96 subjects, computed from the Human Connectome Project \cite{McNab2013}. The user may alternatively choose Version 1.0, that applies only six graphs computed and described by the classical article of \cite{Hagmann2008}.	
	
\item[(ii)] Minimum edge confidence: The graph to be constructed will contain all the edges that are present in at least $k$ graphs, between the very same vertices in each graph. The valid choices for $k=1,\ldots,96$. The last choice means that each source graph needs to contain the edge in order to be presented in the resulting consensus graph.

\end{itemize}

For each edge $\{u,v\}$ , the weight of that edge is a fraction $n/L$, where $n$ is the number of fibers connecting $u$ and $v$, and $L$ is the average length of the fibers. 

\begin{itemize}

\item[(iii)] Minimum edge weight: One may set a slider to a value of minimum weight required. The returned graph will contain edges whose mean or median weights are larger than or equal to this value. The mode of computation (mean or median) can be set by the next option.
 
\item[(iv)] Weight calculation mode: There are two choices: Median or Mean. Choice "Median" means that from the list of weights appearing as the weights of the same edge in different graphs, we use the "central" element, that is, first we sort the weights, and next the element is chosen that separates the upper half of the weights from the lower half of the weights. "Mean" means the arithmetic average of the weights. The default choice is the median, since the median is more robust than the mean: typically extreme large or small strengths have less impact to the median than to the mean.
\end{itemize}

The resulting graph can be downloaded in CSV or GraphML formats, or can readily be visualized on the webpage. The downloaded file-names inherit the parameter-settings as follows: e.g., the Budapest Reference Connectome Version 2.0 is given as the file "{budapest\_connectome\_2.0\_14\_0\_median.csv}", that is, the csv file contains the graph generated by Version 2.0 of the server, with a minimum confidence of 14 (i.e., each edge of the graph is contained in at least 14 input graphs), and the minimum edge weight is 0 and the weights of the edges of the reference graph are computed as the median of the weights of the corresponding edges of the input graphs. 

The format of the CSV file is demonstrated on Table 1.

The number of the common edges in at least n graphs (n=1,2,...,96) are given on Figure 2.

\begin{figure}[h!]
	\centering
	\includegraphics[width=95mm]{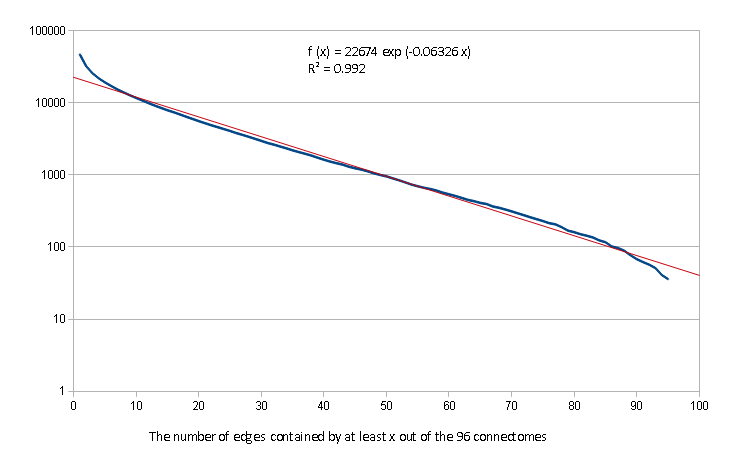}
	\caption{The plot of the number of common edges. }
\end{figure}

\section{Methods} 

The main server, denoted as "v2.0", was created as follows: 

The dataset is a subset of Human Connectome Project 500 Subjects Release (http://www.humanconnectome.org/documentation/S500/), containing MRI images of healthy adult males and females between the ages of 22 and 35. The data was downloaded in October, 2014.

Partitioning, tractography, and graph construction were done by the Connectome Mapper Toolkit (http://cmtk.org).

Partitioning of the grey matter was done by the Lausanne2008 method \cite{Hagmann2008} into 1015 ROIs of surface area of about 1.5 mm$^2$.

For tractography, the deterministic streamline method was applied. 

The graphs were constructed as follows: Two ROIs were connected by an edge if there exists at least one fiber, determined by the tractography step, that connects these two ROIs. The number and the length of the fibers are taken care of by computing the weights of the edges: For each edge $\{u,v\}$ , the weight of that edge is a fraction $n/L$, where $n$ is the number of fibers connecting $u$ and $v$, and $L$ is the average length of the fibers. 

After 96 graphs were computed, each with 1015 vertices, we identified the common edges, their confidence, and weights, computed according to their median and mean. The large, pre-computed tables were integrated into the webserver.

Version 1.0 of the webserver applies the six graphs that were described in \cite{Hagmann2008}. The definition of weight (called strength) and its computation, and also the parcellation of the cortex used are described in \cite{Hagmann2008}.  The six connectomes were downloaded from http://www.cmtk.org/datasets/homo\_sapiens\_01.cff in September, 2014. 

The visualization component applies a modified version of the WebGL Brain Viewer \cite{Ginsburg2011}.

\section{Acknowledgments}
The authors declare no conflicts of interest.



\end{document}